\documentclass[twoside,10pt]{article}
\usepackage{amsmath,amssymb}
\usepackage{xspace}
\usepackage{tam13}

\title{Black holes, TeV-scale gravity and the LHC}

\shorttitle{Black holes and the LHC}

\authors{Elizabeth Winstanley \email{E.Winstanley@sheffield.ac.uk}}

\shortauthors{E.~Winstanley}

\affiliations{Consortium for Fundamental Physics, The University of Sheffield,\\
Hicks Building, Hounsfield Road, Sheffield. S3 7RH United Kingdom}

\abstract{
Over the past 15 years models with large extra space-time dimensions
have been extensively studied.
We have learned from these models that the energy scale of quantum gravity may be
many orders of magnitude smaller than the conventional value of $10^{19}$ GeV.
This raises the tantalizing prospect of probing quantum
gravity effects at the LHC.
Of the possible quantum gravity processes at the LHC, the formation and
subsequent evaporation of microscopic black holes is one of the most spectacular.
We give an overview of some of the fundamental ideas of
the large extra dimensions scenarios and the resulting black hole processes at the LHC.}

\begin{document}
\maketitle

\section{Introduction}

The possibility of producing microscopic black holes at the Large Hadron Collider (LHC)
is one of the most exciting consequences of ``brane world'' models developed over the
past 15 years or so.
The purpose of this presentation is to outline the theories behind this intriguing
possibility:
gravitational theories in which the energy scale of quantum gravity is much lower than
the conventional value of $10^{19}$ GeV, and may possibly be as low as a few TeV.
We will also describe some of the features of black hole creation and evolution
in these low-scale quantum gravity theories.  Finally we link the theoretical modelling
to experimental searches for black holes at the LHC.

There is a vast literature on this subject and in this brief note we cannot do more
than outline a few aspects.  We have not aimed to be complete in our coverage of the
subject, nor in the references.  The reader is encouraged to consult the many
reviews \cite{reviews,kanti} for further details.

\section{Large extra dimensions}

Einstein's theory of general relativity can be formulated in any number of
space-time dimensions,  although all observations to date agree that we live in a
Universe with three space and one time dimension.
Gravitational theories in more than four space-time dimensions have been studied for
many years (beginning with Kaluza-Klein theory in the 1920s).
These extra dimensions have proved invisible to
observations because
their size is smaller than any observable length scale.  For example, in
conventional superstring theory, the extra dimensions are assumed to be compactified
and to have size roughly the order of the Planck length $L_{P}$, the natural
length scale in quantum gravity:
\begin{equation}
L_{P} = {\sqrt {\frac {\hbar G}{c^{3}}}} \sim 10^{-35} \, {\mathrm {m}}.
\label{eq:LP}
\end{equation}

Our focus in this note are higher-dimensional theories with large extra dimensions,
where ``large'' means ``large compared with the Planck length'' (\ref{eq:LP}).
Such theories have been developed as a possible resolution of the hierarchy problem.
The hierarchy problem asks why the natural energy scale of quantum gravity, the Planck
energy $E_{P}$
\begin{equation}
E_{P} = {\sqrt {\frac {\hbar c^{5}}{G}}} \sim 10^{19} \, {\mathrm {GeV}},
\label{eq:EP}
\end{equation}
is so much bigger (about 17 orders of magnitude)
than the natural energy scales of the other fundamental forces in
nature (for example, the electroweak scale is about 100 GeV).
We restrict our attention to one model with large extra dimensions, the ADD scenario
\cite{ADD}.

In the ADD model, there are $n$ extra dimensions
which are
compactified on a length scale $R$.
A comparatively large volume for these extra compact dimensions lowers the
fundamental, higher-dimensional scale of quantum gravity to $E_{*}\ll E_{P}$.
By integrating out the $n$ extra dimensions, the effective scale of quantum gravity
observed in four dimensions, $E_{P}$ (\ref{eq:EP}), is related to $E_{*}$ by
\begin{equation}
E_{P}^{2} \sim R^{n} E_{*}^{2+n}.
\label{eq:Estar}
\end{equation}
By making the size of the extra dimensions sufficiently large, the energy scale
$E_{*}$ can be made many orders of magnitude smaller than $E_{P}$.
For example, for $n=5$ extra dimensions and $R\sim 10^{-13} \, {\mathrm {m}}$,
we have $E_{*}\sim 1 \, {\mathrm {TeV}}$, within the LHC energy range.

While length scales of this size have not been probed gravitationally, they
have been probed using particle physics experiments.
In order to avoid a contradiction with the Standard Model of Particle Physics,
the higher-dimensional space-time in the ADD scenario is comprised of a
four-dimensional brane which is embedded into a higher-dimensional bulk space-time.
All Standard Model particles and forces are confined to the brane, and only
gravitational degrees of freedom can propagate in the bulk.
The effective energy scale for quantum gravity on the brane is $E_{P}$ (\ref{eq:EP}),
while the higher-dimensional energy scale for quantum gravity in the bulk
is $E_{*}$ (\ref{eq:Estar}).
Models such as the ADD scenario are known as ``brane-world'' models.

Theories with large extra dimensions and low-scale quantum gravity, such as the
ADD scenario outlined above, have many interesting consequences.
For example, a collider experiment with centre-of-mass energy
${\sqrt {s}}> E_{*}$ will probe the strong gravity regime.
This raises the exciting possibility that quantum gravity effects may not be many
orders of magnitude out of the reach of terrestrial experiments, but may be probed
at the LHC.
In this note we are concerned with one of the most spectacular
strong gravity processes, namely the
production of black holes in high-energy collisions
\cite{production,dimopoulos,giddingsthomas}.

\section{Production of microscopic black holes}

The basic idea behind the production of black holes in high-energy particle collisions
is very simple.  Consider two colliding
particles whose combined centre-of-mass energy is somewhat larger than $E_{*}$.
In four space-time dimensions, Thorne's ``Hoop Conjecture'' \cite{hoop} states
that a black hole will form if the energy of the particles is compressed into a region
whose circumference in every direction is less than $2\pi r_{H}$, where $r_{H}$ is the
radius of a Schwarzschild black hole having energy equal to the total energy of the
particles.
In more than four space-time dimensions, the ``Hoop Conjecture'' is modified slightly
\cite{hyperhoop}, but the fundamental principle is the same:
if the energy of the colliding
particles is squeezed into a small enough region, then a black hole is expected
to form.

\begin{figure}[h]
\centering
\includegraphics[width=7cm]{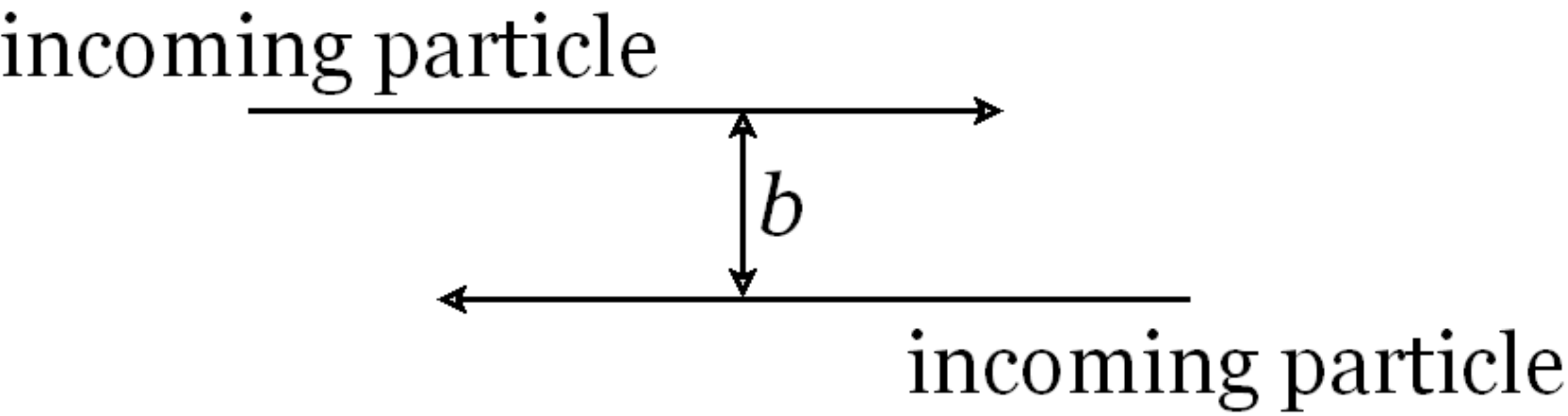}
\caption{Two colliding particles with impact parameter $b$.
\label{impact}}
\end{figure}

For the moment, let us consider the formation of a black hole as a purely classical
process, described by general relativity.
Consider two colliding particles (modelling the partons inside the protons
at the LHC), their distance of closest approach being the impact
parameter $b$ (see Figure~\ref{impact}).
The quantity of prime importance for collider physics is the parton-level production
cross-section $\sigma $, which is required for simulating full production
cross-sections for experimental searches.
The parton-level production cross-section is related to the maximum impact parameter
$b_{max}$ for which a black hole forms from the colliding particles
by the geometric formula
\begin{equation}
\sigma = \pi b_{max}^{2}.
\label{eq:sigma}
\end{equation}
There are two main approaches to studying these collisions and hence finding $b_{max}$.

\begin{figure}[h]
\centering
\includegraphics[width=4cm]{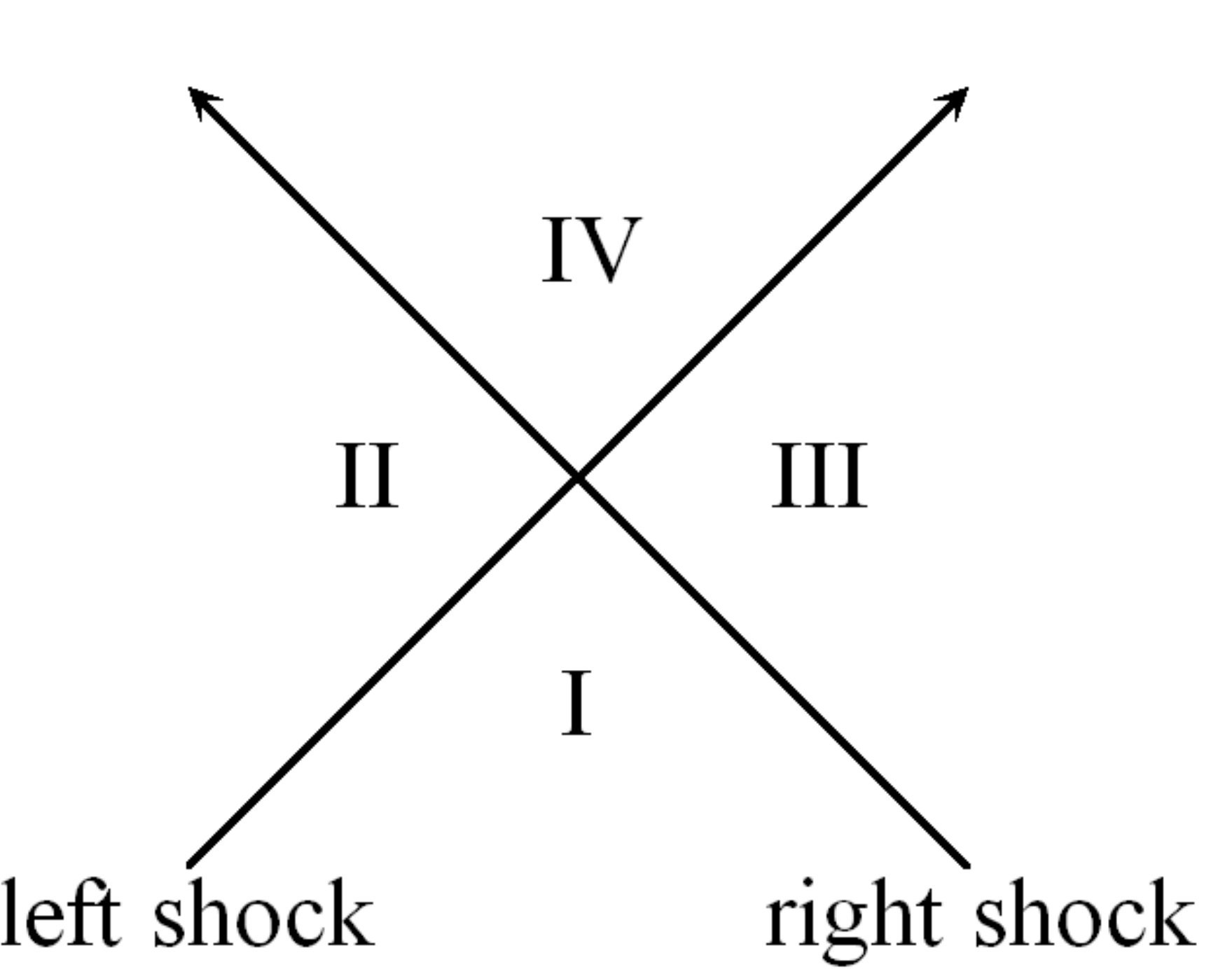}
\caption{The collision of two Aichelburg-Sexl shock waves.
\label{ASshocks}}
\end{figure}

The first models the colliding particles as two gravitational shock waves
\cite{aichelberg}, which are infinitely boosted black holes
(see Figure~\ref{ASshocks}).
The formation of a closed trapped surface in the future of the collision of the two
gravitational shock waves (region IV in Figure \ref{ASshocks})
indicates that a black hole has formed.
The advantage of this approach is that the metric for each gravitational shock wave
is known analytically.
The two shock wave metrics can be superimposed in regions I-III in
Figure~\ref{ASshocks} because, by causality, they can only affect each other in
region IV.
The equation defining a closed trapped surface in region IV has, in general,
to be solved numerically.
In four space-time dimensions, for non-zero
impact parameter $b$, this approach was pioneered by
Eardley and Giddings \cite{eardleygiddings}
building on unpublished work of Penrose and four-dimensional work of D'Eath and Payne
for the $b=0$ case \cite{d'eath}.
Subsequently higher-dimensional collisions have also been investigated
\cite{yr,sampaioAS}.

In the second approach, the colliding particles are modelled
as boson stars, fluid particles or black holes and
the space-time which evolves as the two objects collide
is computed using full numerical relativity.
Over the past eight or so years, numerical relativity has made enormous advances, which
have allowed ultra-relativistic collisions between high-velocity objects to be
studied.
To date, most attention has focussed on collisions in four space-time dimensions
\cite{4dNR}.
Higher-dimensional work is, comparatively, in its infancy (see \cite{hdNR}
for some recent reviews).

In both of the above approaches, the maximum impact parameter $b_{max}$ is found
to be a numerical factor times $r_{H}$, the event horizon radius of a black hole
having the same energy as the combined energy of the colliding particles.
The maximum impact parameter $b_{max}$ then gives the parton-level
geometric production cross-section $\sigma $ (\ref{eq:sigma}).
Some example values of $\sigma $ are given in Table~\ref{sigma} \cite{yr}.
\begin{table}[h]
\centering
\begin{tabular}{|c||c|c|c|c|c|c|c|c|}
\hline
$n$ & 0 & 1 & 2 & 3 & 4 & 5 & 6 & 7
\\
\hline
$\sigma /(\pi r_{H}^{2})$ &
0.71 & 1.54 & 2.15 & 2.52 & 2.77 & 2.95 & 3.09 & 3.20
\\
\hline
\end{tabular}
\caption{Parton-level black hole production cross-sections $\sigma $
for various numbers of space-time dimensions $D=4+n$ \cite{yr}.
\label{sigma}}
\end{table}

The parton-level black hole production cross-sections then feed into the full
production cross-sections \cite{dimopoulos,giddingsthomas}.
There are many parameters involved in estimating these cross-sections
(for example, the value of
$E_{*}$ and the number of space-time dimensions).
Depending on the the values of these parameters, it is possible to construct very
optimistic cross-sections; for example, with $E_{*} = 1$ TeV and
$n=6$ extra dimensions, the production cross-section for black holes with a mass
of 5 TeV/c${}^{2}$ is about one black hole per second (that is, about $10^5$ fb)!
However, the cross-section decreases very rapidly as the black hole mass (or $E_{*}$)
increases; for example, keeping $E_{*} = 1$ TeV and
$n=6$ extra dimensions, the cross-section for black holes with a mass of 10 TeV/c${}^{2}$
is about 10 fb (which is still significant).

We emphasize that the production of black holes at the LHC is only a realistic
possibility for higher-dimensional models outlined in the previous section,
in which the fundamental energy scale of quantum gravity is about $10^{0}-10^{1}$ TeV.
Any black holes which form will be microscopic in scale, having radii about
$\sim 10^{-4}$ fm.

\section{Microscopic black hole decay}

We now consider what happens to a microscopic black hole formed by particle
collisions at the LHC.
When initially created, the black hole will be highly asymmetric and will have
attached gauge field hair arising from the gauge field quantum numbers of the colliding
partons.
The black hole will also be rapidly rotating, due to the initial angular momentum
in the configuration shown in Figure~\ref{impact}.
We assume that the initial energy of the black hole is at least a few times greater
than the quantum gravity scale $E_{*}$, so that its geometry can be described in terms
of general relativity  (see \cite{meade} for more on this assumption).
This is the semi-classical approximation - we consider quantum processes on the
classical black hole background.

The subsequent evolution of the black hole can be described in terms of four
stages \cite{giddingsthomas}:
\begin{description}
\item[{\textit {Balding phase}}]
The black hole sheds its asymmetries and attached gauge field hair.
This phase is often modelled as part of the black hole production process.
At the end of this phase the black hole is still rapidly rotating.
\item[{\textit {Spin-down phase}}]
The black hole emits Hawking radiation, losing mass and angular momentum.
At the end of this phase the black hole is not rotating.
\item[{\textit {Schwarzschild phase}}]
The black hole is now spherically symmetric and continues to emit Hawking radiation.
\item[{\textit {Planck phase}}]
When the energy of the black hole is of the same order as the quantum gravity scale
$E_{*}$, its geometry can no longer be described by general relativity and the full
details of quantum gravity effects (which are ignored in the semi-classical
approximation) become important.
\end{description}
We now briefly discuss each of these phases.

\subsection{Balding phase}

One of the key questions concerning the balding phase is how much
of the initial energy of the colliding particles is shed in gravitational radiation
as the black hole forms.
Both the colliding shock wave model and full numerical relativity calculations outlined
above give upper bounds on this and therefore lower bounds on the mass of the
black hole.
For example, for head-on colliding shock waves, the energy of the black hole
is at least 70\% of the initial energy for collisions in four space-time dimensions,
and at least 58\% of the initial energy for collisions in eleven
space-time dimensions \cite{eardleygiddings}.
As an example of the results from numerical relativity,
four-dimensional calculations indicate that about 50\% of the initial energy of the
colliding particles is radiated away in the ultra-relativistic limit \cite{sper1}.
The emitted gravitational radiation has also been studied by a number of other
approaches - see the reviews \cite{reviews,kanti} for fuller discussions.

The second aspect of the balding phase is the shedding of charges and gauge field hair.
This has not received much attention in the literature.
In particular, QCD effects are likely to be very important at the LHC, but
there is little work on this \cite{QCD}.
The effect of electric charge on the formation process has been studied in numerical
relativity \cite{zilhao}, and upper bounds on the amount of electromagnetic
as well as gravitational radiation have been computed.

Naively one would expect that any electric charge left on the black hole would
rapidly discharge due to Schwinger pair production. However, this assumption is
based on conventional
four-dimensional gravity models where electromagnetic interactions are many
orders of magnitude stronger than gravitational interactions.
In higher-dimensional gravity models with strong gravity, the loss of electric
charge is not so rapid \cite{sampaiocharge}.

\subsection{Black hole at the end of the balding phase}

At the end of the balding phase, the higher-dimensional black hole that remains is
uncharged, axisymmetric and rapidly rotating.
The space of solutions of general relativity describing rotating black objects
in more than four space-time dimensions is extremely rich \cite{hdbhs}
(see \cite{braneBHs} for discussions of black holes in brane world models).
Here we employ a very simple model of the black hole.
Working in the ADD scenario, we assume that the extra dimensions are flat and that
the black hole is very much smaller than the compactification radius of the extra
dimensions (so that the compactification can effectively be ignored).
We also assume that the brane has no tension or energy density.

In higher-dimensional vacuum general relativity, the generalization of the
four-dimensional Kerr geometry describing a rotating black hole is the Myers-Perry
family of metrics \cite{myers}.
For black holes created by particle collisions on the brane,
we are interested in black holes which have a single axis of rotation,
lying in the brane.
In this case the Myers-Perry metric takes the form
\begin{eqnarray}
ds^{2} & = & \left( 1- \frac {\mu }{\Sigma  r^{n-1}} \right) dt^{2}
+ \frac {2a \mu \sin ^{2} \theta }{\Sigma  r^{n-1}} dt \, d\varphi
- \frac {\Sigma }{\Delta _{n}} dr^{2}
- \Sigma \, d\theta ^{2}
\nonumber \\ & &
- \left( r^{2} + a^{2} + \frac {a^{2} \mu \sin ^{2} \theta }{\Sigma r^{n-1}}
\right) \sin ^{2} \theta \, d\varphi ^{2}
- r^{2} \cos ^{2} \theta \, d\Omega _{n}^{2}
\label{eq:MPfull}
\end{eqnarray}
where
\begin{equation}
\Delta _{n} = r^{2} + a^{2} - \frac {\mu }{r^{n-1}}, \qquad
\Sigma = r^{2} + a^{2} \cos ^{2} \theta .
\label{eq:Delta}
\end{equation}
The parameters $\mu $ and $a$ determine the mass $M$ and angular momentum $J$
of the black hole:
\begin{equation}
M = \frac {\left( n+2 \right) A_{n+2}\mu }{16 \pi G_{4+n}}, \qquad
J = \frac {2aM}{n+2}
\end{equation}
where $n$ is the number of extra dimensions, $A_{n+2}$ is the surface area of an
$(n+2)$-dimensional unit sphere and $G_{4+n}$ is the higher-dimensional Newton
constant.
The black hole has an event horizon at $r=r_{H}$, which is the largest positive
root of the equation $\Delta _{n}=0$.
The event horizon rotates with an angular velocity
\begin{equation}
\Omega _{H} = \frac {a}{r_{H}^{2}+a^{2}}.
\label{eq:OmegaH}
\end{equation}
We emphasize that the metric (\ref{eq:MPfull}) is a solution of the $(n+4)$-dimensional
vacuum Einstein equations.

\begin{figure}[h]
\centering
\includegraphics[width=5cm]{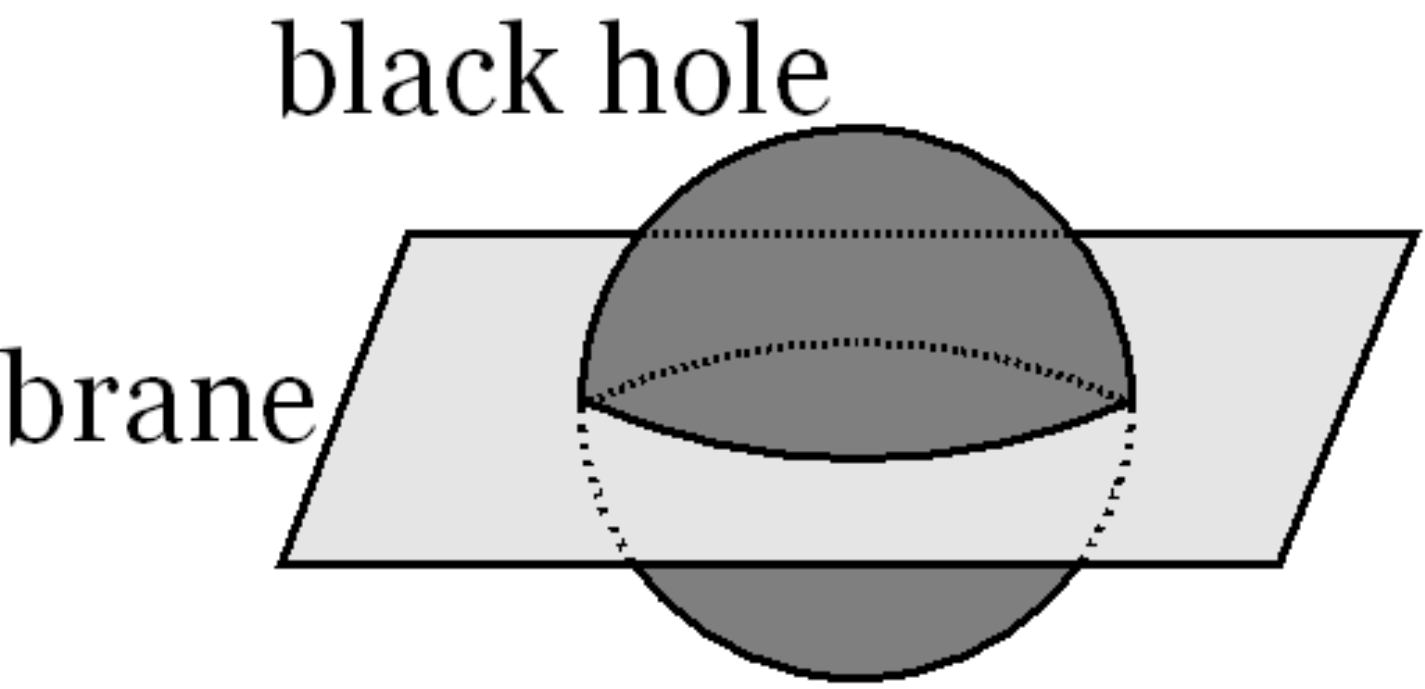}
\caption{The geometry of a higher-dimensional black hole on the brane.
\label{branebh}}
\end{figure}

In the ADD model, the higher-dimensional black hole will lie on the brane (on which
the Standard Model particles and forces are confined), as shown in Figure~\ref{branebh}.
In the full Myers-Perry metric (\ref{eq:MPfull}), the co-ordinates
$(t, r, \theta ,\varphi )$ are the co-ordinates on the brane and
$d\Omega _{n}^{2}$ is the part of the metric coming from the extra dimensions.
Fixing the co-ordinates in the extra dimensions, the metric on the brane ``slice''
of the Myers-Perry black hole takes the form:
\begin{eqnarray}
ds^{2} & = & \left( 1- \frac {\mu }{\Sigma  r^{n-1}} \right) dt^{2}
+ \frac {2a \mu \sin ^{2} \theta }{\Sigma  r^{n-1}} dt \, d\varphi
- \frac {\Sigma }{\Delta _{n}} dr^{2}
- \Sigma \, d\theta ^{2}
\nonumber \\ & &
- \left( r^{2} + a^{2} + \frac {a^{2} \mu \sin ^{2} \theta }{\Sigma r^{n-1}}
\right) \sin ^{2} \theta \, d\varphi ^{2}.
\label{eq:MPbrane}
\end{eqnarray}
This is the four-dimensional black hole metric seen by observers on the brane.
Note that the metric (\ref{eq:MPbrane}) still depends on the number of extra
dimensions $n$.  For $n=0$ it reduces to the usual Kerr metric.
However, for $n>0$ the metric (\ref{eq:MPbrane}) is not a solution of the vacuum
Einstein equations in four space-time dimensions \cite{sampaiocharge}.

\subsection{Spin-down and Schwarzschild phases}

During these two phases of the black hole evolution, the black hole metric is
assumed to be classical.
The black hole will emit quantum thermal Hawking radiation
\cite{hawking}, with a temperature
given by
\begin{equation}
T_{H} = \frac{(n + 1)r_{H}^{2} + (n-1)a^{2}}{4\pi (r_{H}^{2} + a^{2})r_{H}}
\label{eq:temperature}
\end{equation}
for the Myers-Perry black hole (\ref{eq:MPfull}) (the temperature is the same
on the brane and in the bulk space-time).
The semi-classical approximation remains valid as long as the energy of each emitted
quantum is a small proportion of the energy of the black hole.
This remains true until the energy of the black hole is close to the energy scale of
quantum gravity $E_{*}$, when the Planck regime is reached.

In the ADD scenario, Standard Model particles (fermions, gauge bosons and Higgs)
can only be emitted on the brane.
In contrast, gravitons (and possibly some scalars associated with gravitational
degrees of freedom) can be emitted both on the brane and in the bulk.
While the gravitational radiation in the bulk cannot be observed, it contributes
to the missing energy in black hole events.

The Hawking radiation for each species is computed by starting with the
classical field equations on the bulk black hole background (\ref{eq:MPfull})
or brane black hole background (\ref{eq:MPbrane}) as appropriate.
For four-dimensional Kerr black holes, Teukolsky \cite{teukolsky} developed a
formalism which writes the equations for fields of spin $0$ (scalars),
$\frac {1}{2}$ (fermions), $1$ (gauge bosons) and $2$ (gravitons) in terms of a single
``master'' equation for a quantity $\Psi $ (the exact nature of $\Psi $ depending on
the spin of the field).
Teukolsky's formalism can be extended to fields of spin $0$, $\frac {1}{2}$ and $1$
(that is,
the fields of the Standard Model) on the brane metric (\ref{eq:MPbrane}) (see
\cite{kanti,brane} for more details).
The field quantity $\Psi $ is then expanded in terms of modes of frequency $\omega $:
\begin{equation}
\Psi =
\sum _{\omega \ell m} R_{s \omega \ell m}(r) S_{s \omega \ell m}(\theta )
e^{-i\omega t} e^{im\varphi },
\end{equation}
where $s$ is the spin of the field, $\ell $ the total angular momentum quantum number
and $m$ the azimuthal quantum number.
The angular function $S_{s\omega \ell m} (\theta )$ is a spin-weighted spheroidal
harmonic, and the radial function $R_{s\omega \ell m}(r)$ can only be computed
numerically.

To study the Hawking radiation, we are interested in ``up'' modes for which
the radial function takes the typical form
\begin{equation}
R_{s\omega \ell m} = \left\{
\begin{array}{ll}
e^{i {\tilde {\omega }} r_{*}} + A_{\omega \ell m}^{{\mathrm {up}}}
e^{-i{\tilde {\omega }} r_{*}},
& r_{*}\rightarrow -\infty ,
\\
B_{\omega \ell m}^{{\mathrm {up}}} e^{i\omega r_{*}},
& r_{*} \rightarrow \infty .
\end{array}
\right.
\label{eq:up}
\end{equation}
In the above equation, we have written the radial function $R_{s\omega \ell m}$
as a function of the ``tortoise'' co-ordinate $r_{*}$, which is defined in terms of the
co-ordinate $r$ by
\begin{equation}
\frac {dr_{*}}{dr} = \frac {r^{2}+a^{2}}{\Delta _{n}}
\end{equation}
for the metrics (\ref{eq:MPfull}, \ref{eq:MPbrane}), where $\Delta _{n}$ is given by
(\ref{eq:Delta}).
The frequency of the mode as seen by an observer far from the black hole is
$\omega $, but due to the rotation of the black hole an observer near the
event horizon sees a modified frequency
\begin{equation}
{\tilde {\omega }}=\omega - m\Omega _{H},
\end{equation}
where $\Omega _{H}$ is the angular velocity of the event horizon (\ref{eq:OmegaH}).
In (\ref{eq:up}), $A_{\omega \ell m}^{{\mathrm {up}}}$ and
$B_{\omega \ell m}^{{\mathrm {up}}}$ are complex constants for each mode.
The ``up'' modes (\ref{eq:up}) represent waves which emanate from the event horizon
of the black hole.
Part of the wave (the part involving $A_{\omega \ell m}^{{\mathrm {up}}}$)
is reflected back down the black hole, and part of the wave (the part involving
$B_{\omega \ell m}^{{\mathrm {up}}}$) is transmitted out to infinity.
It is the latter part which contributes to the Hawking radiation observed far
from the black hole.

For each particle species, the Hawking radiation is computed by summing
the contributions from each ``up'' mode.  The differential emission rates
per unit time and unit frequency $\omega $ of particles ($N$), energy ($E$)
and angular momentum ($J$) are given by
\begin{equation}
\frac {d^{2}}{dt \, d\omega }
\left(
\begin{array}{c}
N \\ E \\ J
\end{array}
\right)
= \frac {1}{2\pi } \sum _{{\rm {modes}}}
\frac {\left| {\cal {A}}_{s \omega \ell m} \right| ^{2}}{
e^{{\tilde {\omega }}/T_{H}}\mp 1}
\left(
\begin{array}{c}
1 \\ \omega \\ m
\end{array}
\right)
\label{eq:fluxes}
\end{equation}
where $T_{H}$ is the Hawking temperature (\ref{eq:temperature}), the $+$ sign
is for fermions, the $-$ sign for bosons, and we have integrated over all angles.
The quantity $\left| {\cal {A}}_{s \omega \ell m} \right| ^{2}$
is known as the grey-body factor.
It encodes the fact that the emitted radiation is not exactly thermal, due to the
interaction of the emitted quanta with the gravitational potential which
surrounds the black hole.
The grey-body factor corresponds to the proportion of the flux emitted
near the event horizon of the black hole which tunnels through the potential
barrier to reach infinity.
It is computed from the ``up'' modes (\ref{eq:up}) as follows:
\begin{equation}
\left| {\cal {A}}_{s \omega \ell m } \right| ^{2}
= 1- \left| A_{\omega \ell m}^{{\mathrm {up}}} \right| ^{2}.
\end{equation}

There is a considerable body of work studying the Hawking radiation in these two
phases of the evolution of the black hole.
For the sake of brevity, here we discuss only results for neutral, massless fields
and only give a very limited number of references.
Fuller discussions and more complete lists of references can be found in the detailed
reviews \cite{reviews,kanti}.
A summary of what is known about the Hawking radiation in the spin-down and
Schwarzschild phases is presented in Table~\ref{radiation}.

\begin{table}[h]
\centering
\begin{tabular}{| l | l |}
\hline
Spin-down phase & Schwarzschild phase
\\
\hline
\hline
{\textit {Brane emission}} \cite{brane}
&
{\textit {Brane emission}} \cite{schwbrane}
\\
scalars, fermions, gauge bosons
&
scalars, fermions, gauge bosons
\\
\hline
{\textit {Bulk emission}} \cite{scalars}
&
{\textit {Bulk emission}} \cite{schwbrane}
\\
scalars
&
scalars
\\
\hline
{\textit {Graviton emission}} \cite{spingravitons}
&
{\textit {Graviton emission}} \cite{schwgravitons,cardoso}
\\
partial results only
&
complete results
\\
\hline
\end{tabular}
\caption{
Summary of Hawking radiation results during the spin-down and Schwarzschild phases of the
evolution of the black hole, for neutral massless particles only.
\label{radiation}}
\end{table}

The radiation of Standard Model particles on the brane in both the spin-down
\cite{brane} and Schwarzschild phases \cite{schwbrane} is studied using
a  generalization of Teukolsky's formalism \cite{teukolsky}.
Due to the comparative simplicity of the scalar field equation
(even on the higher-dimensional metric (\ref{eq:MPfull}))
the radiation of scalar particles on the brane and
in the bulk is tractable \cite{brane,schwbrane,scalars}.
Teukolsky's original formalism \cite{teukolsky} was applicable to graviton
(spin-2) perturbations of four-dimensional Kerr black holes, but does not easily
 generalize
to gravitational perturbations of higher-dimensional black holes.
The formalism for dealing with gravitational perturbations of spherically symmetric
higher-dimensional black holes has been developed \cite{kodama}, which
has enabled the Hawking radiation to be studied in this case \cite{schwgravitons,cardoso}.
However, for higher-dimensional rotating black holes, the
perturbation equations are much more complicated, even for singly-rotating
Myers-Perry black holes (\ref{eq:MPfull}) \cite{mpperts}.
Unlike the spherically symmetric case, the equations describing gravitational
perturbations of Myers-Perry black holes do not separate into ordinary differential
equations \cite{mpperts}, which has rendered computing the Hawking radiation
intractable to date.
The exception to this is tensor-type gravitational perturbations
of Myers-Perry black holes, which satisfy the separable
scalar field equation.
There are results for the Hawking radiation for this restricted class of
gravitational perturbations \cite{spingravitons}.

A key question for experimental searches is how much of the Hawking radiation
is emitted on the brane (since only radiation on the brane is observable).
Due to the large number of degrees of freedom in the Standard Model, and the
democratic emission of Hawking radiation, it is expected that most radiation will be
on the brane \cite{brane-bulk}.
However, the number of gravitational degrees of freedom increases
rapidly with increasing $n$ (the number of extra dimensions).
This means that the proportion of the Hawking radiation escaping into the bulk also
increases rapidly as $n$ increases (see Table~\ref{grav}), although, even
in eleven space-time dimensions ($n=7$), three-quarters of the radiation is on the brane.
\begin{table}[h]
\centering
\begin{tabular}{|c||c|c|c|c|c|}
\hline
& $\,\,\, n=0 \,\,\, $ & $\,\,\, n=1\,\,\, $ &
 $\,\,\, n=3 \,\,\, $ &
 $\,\,\, n=5 \,\,\, $ &
 $\,\,\, n=7 \,\,\, $ \\
 \hline
 scalars &
 6.8 & 4.0  & 3.6  & 3.5  & 2.9
 \\
 \hline
 fermions &
 83.8 & 78.7  & 72.3 & 68.1  & 53.4
 \\
 \hline
 gauge bosons &
 9.3 & 16.7 &  21.7 &  22.2  & 18.6
 \\
 \hline
 gravitons  &
 0.1 & 0.6  & 2.4  & 7.7  & 25.1
 \\
 \hline
\end{tabular}
\caption{Percentages of energy emission from a non-rotating black hole
into particles of spin-$0$, $\frac {1}{2}$,
$1$ and $2$, assuming the Standard Model of particle physics with three families and
one Higgs field on the brane and only graviton emission in the bulk.
Data taken from \cite{cardoso}.}
\label{grav}
\end{table}

\section{Quantum black holes}

Our focus in this brief note has been the ``standard'' model of microscopic
black hole production and decay at the LHC, in the context of the ADD brane-world scenario.
In this model, the black hole is semi-classical: the metric is classical and
described by general relativity, and the black hole emits quantum Hawking radiation.
This semi-classical approximation breaks down when the black hole energy is roughly
$E_{*}$, the energy scale at which the details of the unknown theory of quantum
gravity become important.
Meade and Randall \cite{meade} have argued that, in order for the black hole to
be described by a classical metric, it must be the case that the Compton wavelengths of
the colliding particles lie within the event horizon of the formed black hole.
This implies that the black hole energy should be at least an order of magnitude larger
than $E_{*}$ for the semi-classical approximation to be valid.

In the absence of a full theory of quantum gravity, there have been attempts in the
literature to study fully quantum black holes with energies close to $E_{*}$
\cite{QCD,quantum}, as well to refine the semi-classical picture to incorporate
quantum gravity effects \cite{nicolini}.
Fully quantum black holes do not decay thermally, but instead emit just a few particles.
Particle physics symmetries are used to constrain the decay processes.

\section{Experimental searches}

There are a number of event generators simulating black hole processes at the LHC
\cite{charybdis,blackmax,generators,QBH}.
The LHC experimental groups use CHARYBDIS2 \cite{charybdis} and
BlackMax \cite{blackmax} for simulating semi-classical black holes
and QBH \cite{QBH} for the simulation of quantum black holes.
Black hole events typically have high primary particle multiplicity
with large missing transverse momentum.

At the time of writing no evidence for either semi-classical or quantum black holes
has been observed at the LHC \cite{ATLAS,CMS}.
These null results have enabled the LHC experimental groups to set lower bounds on
the higher-dimensional quantum gravity scale $E_{*}$.
ATLAS rule out semi-classical black holes having masses lower than about
4 TeV/c${}^{2}$ for six extra dimensions and $E_{*}$ about 2 TeV \cite{ATLAS}, while
CMS have slightly higher lower bounds on the semi-classical black hole mass
for the same
values of $E_{*}$ \cite{CMS}.
CMS also rule out quantum black holes with masses lower than
about $5-6$ TeV/c${}^{2}$ for $E_{*}=2-5$ TeV \cite{CMS}.

\section{Conclusions}

We have briefly reviewed the ADD large extra dimensions scenario, in which the
energy scale of quantum gravity, $E_{*}$, may be as low as a few TeV.
This raises the exciting possibility of probing quantum gravity effects at the LHC.
Of the many possible strong gravity processes, those involving microscopic black holes
will be some of the most spectacular.
Our focus in this note has been a semi-classical model of microscopic black hole
production and decay, with the geometry described by general relativity and
quantum Hawking radiation being emitted from the black hole.
We have also discussed the validity of this model and recent work on describing
fully quantum black holes.
To date, there has been no experimental evidence for black holes at the LHC.
However, this does not diminish the importance of searching for them: the
non-observation results have set lower bounds on the energy scale $E_{*}$,
constraining the elusive theory of quantum gravity.

\section*{Acknowledgements}
{\footnotesize
EW thanks the organizers of TAM 2013 for a very enjoyable and stimulating conference in
one of the most beautiful cities in the world.
She also thanks her collaborators on work relevant to this talk:
Marc Casals, Sam Dolan, Gavin Duffy, Chris Harris, Panagiota Kanti and
Piero Nicolini. This work is supported by
the Lancaster-Manchester-Sheffield Consortium for Fundamental Physics under
STFC Grant No. ST/J000418/1 and by EU COST Action MP0905
``Black Holes in a Violent Universe''.
}


\begin{thebibliography}{99}
\raggedright
\footnotesize

\bibitem{reviews}
A.~Casanova and E.~Spallucci, Class.\ Quantum Grav.\ {\textbf {23}} (2006) R45;
M.~Cavagli\`a, Int.\ J.\ Mod.\ Phys.\ A {\textbf {19}} (2003) 1843;
S.~Hossenfelder, in {\textit {Focus on black hole research}}, ed.\ P.V.\ Kreitler,
pp.\ 155--192
(Nova Science Publishers, 2005);
P.~Kanti, Lect.\ Notes Phys.\ {\textbf {769}} (2009) 387;
P.~Kanti, Rom.\ J.\ Phys.\ {\textbf {57}} (2012) 879;
G.~Landsberg, Eur.\ Phys.\ J.\ C {\textbf {33}} (2004) S927;
A.S.~Majumdar and N.~Mukherjee, Int.\ J.\ Mod.\ Phys.\ D {\textbf {14}}
(2005) 1095;
S.C.~Park, Prog.\ Part.\ Nucl.\ Phys.\ {\textbf {67}} (2012) 617;
B.~Webber, {\tt {hep-ph/0511128}};
E.~Winstanley, {\tt {arXiv:0708.2656}}.

\bibitem{kanti}
P.~Kanti, Int.\ J.\ Mod.\ Phys.\ A {\textbf {19}} (2004) 4899.

\bibitem{ADD}
I.~Antoniadis, N.~Arkani-Hamed, S.~Dimopoulos and G.R.~Dvali,
Phys.\ Lett.\ B {\textbf {436}} (1998) 257;
N.~Arkani-Hamed, S.~Dimopoulos and G.R.~Dvali,
Phys.\ Rev.\ D {\textbf {59}} (1999) 086004.

\bibitem{production}
T.~Banks and W.~Fischler, {\tt {hep-th/9906038}}.

\bibitem{dimopoulos}
S.~Dimopoulos and G.~Landsberg, Phys.\ Rev.\ Lett.\ {\textbf {87}}
(2001) 161602.

\bibitem{giddingsthomas}
S.B.~Giddings and S.~Thomas, Phys.\ Rev. D {\textbf {65}} (2002)
056010.

\bibitem{hoop}
K.S.~Thorne, in {\it {Magic without magic: John Archibald Wheeler. A collection
of essays in honor of his sixtieth birthday}}, ed.\ J.\ Klauder (W.H.\ Freeman,
San Francisco, 1972).

\bibitem{hyperhoop}
D.~Ida and K.-i.~Nakao, Phys.\ Rev.\ D {\textbf {66}} (2002) 064026;
C.m.~Yoo, H.~Ishihara, M.~Kimura and S.~Tanzawa,
  Phys.\ Rev.\ D {\textbf {81}} (2010) 024020.

\bibitem{aichelberg}
P.C.~Aichelburg and R.U.~Sexl,
Gen.\ Rel.\ Grav.\  {\textbf {2}} (1971) 303.

\bibitem{eardleygiddings}
D.M.~Eardley and S.B.~Giddings,
Phys.\ Rev.\ D {\textbf {66}} (2002) 044011.

\bibitem{d'eath}
P.D.~D'Eath, {\textit {Black holes: gravitational interactions}},
(Oxford Science Publications 1996).

\bibitem{yr}
H.~Yoshino and V.S.~Rychkov,
Phys.\ Rev.\ D {\textbf {71}} (2005) 104028
[Erratum-ibid.\ D {\textbf {77}} (2008) 089905].

\bibitem{sampaioAS}
F.~S.~Coelho, C.~Herdeiro and M.~O.~P.~Sampaio,
Phys.\ Rev.\ Lett.\  {\textbf {108}} (2012) 181102;
C.~Herdeiro, M.~O.~P.~Sampaio and C.~Rebelo,
JHEP {\textbf {1107}} (2011) 121;
M.O.P.~Sampaio, {\tt {arXiv:1306.0903}}.

\bibitem{4dNR}
M.W.~Choptuik and F.~Pretorius,
Phys.\ Rev.\ Lett.\  {\textbf {104}} (2010) 111101;
W.E.~East and F.~Pretorius,
Phys.\ Rev.\ Lett.\  {\textbf {110}} (2013) 101101;
L.~Rezzolla and K.~Takami,
Class.\ Quant.\ Grav.\  {\textbf {30}} (2013) 012001;
M.~Shibata, H.~Okawa and T.~Yamamoto,
Phys.\ Rev.\ D {\bf 78} (2008) 101501;
U.~Sperhake, V.~Cardoso, F.~Pretorius, E.~Berti and J.A.~Gonzalez,
Phys.\ Rev.\ Lett.\  {\textbf {101}} (2008) 161101.

\bibitem{hdNR}
U.~Sperhake,
Int.\ J.\ Mod.\ Phys.\ D {\textbf {22}} (2013) 1330005;
H.M.S.~Yoshino and M.~Shibata,
Prog.\ Theor.\ Phys.\ Suppl.\  {\textbf {189}} (2011) 269;
H.M.S.~Yoshino and M.~Shibata,
Prog.\ Theor.\ Phys.\ Suppl.\  {\textbf {190}} (2011) 282.

\bibitem{meade}
P.~Meade and L.~Randall,
JHEP {\textbf {0805}} (2008) 003.

\bibitem{sper1}
U.~Sperhake, E.~Berti, V.~Cardoso and F.~Pretorius,
{\tt {arXiv:1211.6114}}.

\bibitem{QCD}
X.~Calmet, W.~Gong and S.D.H.~Hsu,
Phys.\ Lett.\ B {\textbf {668}} (2008) 20;
D.M.~Gingrich,
J.\ Phys.\ G {\textbf {37}} (2010) 105008.

\bibitem{zilhao}
M.~Zilhao, V.~Cardoso, C.~Herdeiro, L.~Lehner and U.~Sperhake,
Phys.\ Rev.\ D {\textbf {85}} (2012) 124062.

\bibitem{sampaiocharge}
M.O.P.~Sampaio,
JHEP {\textbf {0910}} (2009) 008.

\bibitem{hdbhs}
R.~Emparan and H.S.~Reall,
Living Rev.\ Rel.\  {\textbf {11}} (2008) 6;
S.~Tomizawa and H.~Ishihara,
Prog.\ Theor.\ Phys.\ Suppl.\  {\textbf {189}} (2011) 7.

\bibitem{braneBHs}
R.~Gregory,
Lect.\ Notes Phys.\  {\textbf {769}} (2009) 259;
P.~Kanti,
J.\ Phys.\ Conf.\ Ser.\  {\textbf {189}} (2009) 012020;
N.~Tanahashi and T.~Tanaka,
Prog.\ Theor.\ Phys.\ Suppl.\  {\textbf {189}} (2011) 227.

\bibitem{myers}
R.C.~Myers and M.J.~Perry, Annals Phys.\ {\textbf {172}} (1986) 304.

\bibitem{hawking}
S.W.~Hawking, Commun.\ Math.\ Phys.\ {\textbf {43}} (1975) 199.

\bibitem{teukolsky}
S.A.~Teukolsky, Phys.\ Rev.\ Lett.\ {\textbf {29}} (1972) 1114;
S.A.~Teukolsky, Astrophys.\ J.\ {\textbf {185}} (1973) 635.

\bibitem{brane}
M.~Casals, S.R.~Dolan, P.~Kanti and E.~Winstanley,
JHEP {\textbf {0703}} (2007) 019;
M.~Casals, P.~Kanti and E.~Winstanley,
JHEP {\textbf {0602}} (2006) 051;
G.~Duffy, C.~Harris, P.~Kanti and E.~Winstanley,
JHEP {\textbf {0509}}  (2005) 049;
D.~Ida, K.-y.~Oda and S.C.~Park,
Phys.\ Rev.\ D {\textbf {67}} (2003) 064025
[Erratum-ibid.\ D {\bf 69} (2004) 049901].

\bibitem{schwbrane}
C.M.~Harris and P.~Kanti,
JHEP {\textbf {0310}} (2003) 014.

\bibitem{scalars}
M.~Casals, S.R.~Dolan, P.~Kanti and E.~Winstanley,
JHEP {\textbf {0806}} (2008) 071.

\bibitem{spingravitons}
J.~Doukas, H.T.~Cho, A.S.~Cornell and W.~Naylor,
Phys.\ Rev.\ D {\textbf {80}} (2009) 045021;
P.~Kanti, H.~Kodama, R.A.~Konoplya, N.~Pappas and A.~Zhidenko,
Phys.\ Rev.\ D {\textbf {80}} (2009) 084016.

\bibitem{schwgravitons}
A.S.~Cornell, W.~Naylor and M.~Sasaki,
JHEP {\textbf {0602}} (2006) 012;
S.~Creek, O.~Efthimiou, P.~Kanti and K.~Tamvakis,
Phys.\ Lett.\ B {\textbf {635}} (2006) 39;
D.K.~Park,
Phys.\ Lett.\ B {\textbf {638}} (2006) 246.

\bibitem{cardoso}
V.~Cardoso, M.~Cavagli\`a and L.~Gualtieri,
JHEP {\textbf {0602}} (2006) 021.

\bibitem{kodama}
H.~Kodama and A.~Ishibashi,
Prog.\ Theor.\ Phys.\  {\textbf {110}} (2003) 701.

\bibitem{mpperts}
M.~Durkee and H.S.~Reall,
Class.\ Quant.\ Grav.\  {\textbf {28}} (2011) 035011;
K.~Murata,
Prog.\ Theor.\ Phys.\ Suppl.\  {\textbf {189}} (2011) 210;
H.S.~Reall,
Int.\ J.\ Mod.\ Phys.\ D {\textbf {21}} (2012) 1230001.

\bibitem{brane-bulk}
R.~Emparan, G.T.~Horowitz and R.C.~Myers,
Phys.\ Rev.\ Lett.\  {\textbf {85}} (2000) 499.

\bibitem{quantum}
X.~Calmet, D.~Fragkakis and N.~Gausmann,
chapter 8 in {\textit {Black holes: evolution, theory and thermodynamics}},
ed.\ A.J.~Bauer and D.G.~Eiffel
(Nova Science Publishers, 2012);
X.~Calmet and N.~Gausmann,
Int.\ J.\ Mod.\ Phys.\ A {\textbf {28}} (2013) 135004.

\bibitem{nicolini}
P.~Nicolini and E.~Winstanley,
JHEP {\textbf {1111}} (2011) 075.


\bibitem{charybdis}
J.A.~Frost, J.R.~Gaunt, M.O.P.~Sampaio, M.~Casals, S.R.~Dolan,
M.A.~Parker and B.R.~Webber,
JHEP {\textbf {0910}} (2009) 014.

\bibitem{blackmax}
D.-C.~Dai, G.~Starkman, D.~Stojkovic, C.~Issever, E.~Rizvi and J.~Tseng,
Phys.\ Rev.\ D {\textbf {77}} (2008) 076007.

\bibitem{generators}
M.~Cavagli\`a, R.~Godang, L.~Cremaldi and D.~Summers,
Comput.\ Phys.\ Commun.\  {\textbf {177}} (2007) 506;
D.M.~Gingrich, {\tt {hep-ph/0610219}};
G.L.~Landsberg, J.~Phys.~G {\textbf {32}} (2006) R337.

\bibitem{QBH}
D.M.~Gingrich,
Comput.\ Phys.\ Commun.\  {\textbf {181}} (2010) 1917.

\bibitem{ATLAS}
G.~Aad {\textit {et al.\ }}  [ATLAS Collaboration],
Phys.\ Lett.\ B {\textbf {716}} (2012) 122.

\bibitem{CMS}
S.~Chatrchyan {\textit {et al.\ }}  [CMS Collaboration],
{\tt {arXiv:1303.5338}}.

\end{thebibliography}
\end{document}